\documentstyle[12pt,epsf]{article}
\newcommand{\beq}{\begin{equation}}
\newcommand{\eeq}{\end{equation}}
\newcommand{\beqa}{\begin{eqnarray}}
\newcommand{\eeqa}{\end{eqnarray}}

\begin{document}

%%{\bf DRAFT} 

\hfill KFA-IKP(Th)-1996-11 

\hfill TK 96 26

\hfill nucl-th/9609051

\bigskip\bigskip\bigskip

\begin{center}

{{\Large\bf Threshold kaon photo- and electroproduction in SU(3)
    baryon chiral perturbation theory}}

\end{center}

\vspace{.4in}

\begin{center}
{\large S. Steininger$^\dagger$\footnote{email: sven@pythia.itkp.uni-bonn.de},
 Ulf-G. Mei{\ss}ner$^\ddagger$\footnote{email: Ulf-G.Meissner@kfa-juelich.de}}

\bigskip

\bigskip

$^\dagger$Universit\"at Bonn, Institut f{\"u}r Theoretische Kernphysik\\
Nussallee 14-16, D-53115 Bonn, Germany\\

\bigskip

$^\ddagger$Forschungszentrum J\"ulich, Institut f\"ur Kernphysik (Theorie)\\ 
D-52425 J\"ulich, Germany

\end{center}

\vspace{.7in}

\thispagestyle{empty} 

\begin{abstract}
We calculate the processes $\gamma p \to K^0 \Sigma^+ , K^+ \Sigma^0,
K^+ \Lambda$ in three flavor heavy baryon chiral perturbation theory
to one loop. Some low--energy constants are fixed from single baryon
properties while others are estimated by means of resonance saturation. 
The total cross sections are comparable with the few existing data
points in the threshold region. The angular dependences of the recoil 
polarization of the  $\Lambda$ and $\Sigma^0$ show most features of
the ones measured at ELSA. We also predict the isovector charge radius
of the $\Sigma^+$, $\langle r^2_V \rangle_{\Sigma^+} = (0.30 \pm 0.03)
 \,$fm$^2$.
\end{abstract}

\vfill

\newpage

\noindent {\bf 1.} Threshold pion photo- and electroproduction has
been at the center of numerous experimental studies at MAMI, SAL and
NIKHEF over the last few years. These data serve as one of the major
testing grounds of baryon chiral perturbation theory which allows to 
sharpen our understanding of the spontaneous and explicit chiral
symmetry breaking in QCD. One particularly spectacular example is the
process $\gamma p \to \pi^0 p$ which clearly shows the relevance of 
chiral (pion) loops in the electric dipole amplitude and has also lead
to novel low--energy theorems related to some of the P--wave
multipoles \cite{bkmz}\cite{bkmpi0}. With the emergence of kaon 
photoproduction data from ELSA
%from the SAPHIR detector at ELSA 
at Bonn (total and differential cross
sections as well as hyperon recoil polarizations) \cite{bock}\cite{schwille},
it appears natural to extend the successful SU(2) pion production
calculations to the three--flavor case.\footnote{The data of 
ref.\cite{schwille} are not yet available.}
Evidently, while for SU(2) the
pertinent expansion parameter is small, $M_\pi / 4 \pi F_\pi = 0.12$
(with $M_\pi$ and $F_\pi$ the pion mass and decay constant, respectively),
the larger strange quark mass leads to $M_K / 4 \pi F_\pi = 0.43$ 
(with $M_K$ the kaon mass). Therefore, it is 
a priori  not clear whether the method of expanding
S--matrix elements and transition currents in small momenta and meson
masses, which is at the heart of chiral perturbation theory (CHPT), 
is applicable in the presence of baryons. 
In this letter, we show the results of such an exploratory study which
lends some credit to the usefulness of the method.
 
\bigskip

\noindent {\bf 2.} The starting point of the heavy baryon 
CHPT \cite{jm}\cite{bkkm} is an
effective Lagrangian formulated in terms of the asymptotic fields,
here the octet of Goldstone bosons (collected in the matrix--valued
field $U(x)$) and the ground state baryon octet, denoted $B$. It
admits a low energy expansion of the form
\beq
{\cal L}_{\rm eff} = {\cal L}_M + {\cal L}_{MB} =
{\cal L}_M^{(2)} +  {\cal L}_{MB}^{(1)} + {\cal L}_{MB}^{(2)}
+ {\cal L}_{MB}^{(3)} + \ldots \,
\eeq  
where the subscript $'M'$ ($'MB'$) denotes the meson (meson--baryon)
sector and the superscript $'(i)'$ the chiral dimension. The ellipsis
stands for terms of higher order not needed here. Beyond leading
order, the effective Lagrangian contains parameters not fixed by
chiral symmetry, the so--called low--energy constants. In principle,
these LECs should be pinned down from data or calculated by means of
lattice gauge theory. To some extent,
their values reflect the spectrum of QCD \cite{reso} which has lead to
the resonance exchange saturation hypothesis. This somewhat
model--dependent method to estimate the LECs is an indispensible
tool in the absence of a sufficiently complete body of accurate data. The
form of ${\cal L}_{MB}^{(1)}$ is standard, we use here the notation of
ref.\cite{bkmzm}, and from ${\cal L}_{MB}^{(2)}$ we only need the
magnetic photon baryon (transition) couplings. At third order in the
energy expansion, we have in total 15 terms contributing to the S-- and
P--waves,
\beq
{\cal L}_{MB}^{(3)} = \sum_{i=1}^{13} d_i \, O_i + 
d_{14} \, O_{14} + d_{15} \, {O}_{15} \,\,\, ,
\eeq
with the last  two terms related to baryon octet isovector charge
radii and only of relevance for electroproduction. The LECs $d_1$ and
$d_2$ enter the axial (transition) radii and one combination can be
fixed from the nucleon axial radius (see below). We work to order
$q^3$ in the small momentum expansion, i.e. we have to consider tree
graphs with insertions from ${\cal L}_{MB}^{(1,2,3)}$ and one loop 
diagrams with insertions from ${\cal L}_{MB}^{(1)}$.

\bigskip

\noindent {\bf 3.} Consider now the reactions $\gamma (k)+ p (p_1) \to
K^0 (q) + \Sigma^+ (p_2)\, , \,  K^+ (q)+ \Sigma^0 (p_2) \, , \, 
K^+ (q) + \Lambda (p_2)$, with $\gamma$ a real ($k^2=0$) or a virtual ($k^2
<0$) photon. In the threshold region, the transition matrix--element can be
written in terms of S-- and P--wave multipoles (for photoproduction),
\beq
\frac{m}{4\pi\sqrt{s}} T\cdot \varepsilon = i {\vec \sigma} \cdot
\vec{\varepsilon} \, ( E_{0+} + \hat{k}\cdot \hat{q} \, P_1 ) + i  {\vec
\sigma} \cdot \hat{k} \, \vec{\varepsilon} \cdot \hat{q} \, P_2 + (\hat{q}
\times \hat{k}) \cdot \vec{\varepsilon} \, P_3 \,\, ,
\eeq 
with $m$ the proton mass, $s =(p_1+k)^2$ the cms energy squared
and $\varepsilon_\mu$ the photon polarization vector. 
These four amplitudes (and the additional $L_{0+}, P_{4,5}$ for 
electroproduction) are calculable within CHPT. Note that
we do not expand the prefactor $m/\sqrt{s}$ in what
follows. Generically, the chiral expansion of the various multipoles
takes the form
\beq
{\cal M} = {\cal M}^{\rm Born} + {\cal M}^{\rm loop} + {\cal M}^{\rm
  ct} \,\,\, ,
\eeq
where the Born terms subsume the tree graphs and the anomalous magnetic
moment couplings from ${\cal L}_{MB}^{(2)}$,
'ct' refers to the remaining counter terms and the
loop contribution at order $q^3$ is given entirely in terms of the
parameters from ${\cal L}_M^{(2)} + {\cal L}_{MB}^{(1)}$. These are
the Goldstone boson decay constant $F_\phi$ (we choose here $F_\phi
= F_K = 113\,$MeV) and the two axial couplings $F \simeq0.5$ and 
$D \simeq0.75$.
The magnetic moments appearing in the Born terms are all well known
with the exception of the one for the $\Sigma^0$. To determine it, we
use the Coleman--Glashow relation (which is exact to the order we are
working), $\mu_{\Sigma^0} = [ \mu_{\Sigma^+} + \mu_{\Sigma^-} ] /2  
= 0.65 \,$n.m.$\,$. For the coupling constants, 
we use the SU(3) predictions
$\sqrt{6} g_{p\Lambda K^+} = -(D+3F) g_{\pi N} / (F+D)$ and 
$\sqrt{2} g_{p\Sigma^0 K^+} = (D-F) g_{\pi N}/(F+D)$ with $g_{\pi N}
= 13.4$ the pion--nucleon coupling constant.
Clearly, a more precise knowledge of these coupling
constants would be needed to reduce the uncertainty from the Born
graphs. For charged kaon photoproduction, one combination of
the LECs $d_1$ and $d_2$ can be inferred from the nucleon axial
radius, $\langle r_A^2 \rangle = 0.42\,$fm$^2$. 
This constrains the $p \to \Lambda\, K^+$ transition axial radius,
\beq
|\langle r_A^2 \rangle_{p \to \Lambda \, K^+}| = \frac{3
  \sqrt{2}}{D+3F} (\, d_1 + 3 d_2 \,) 
= 0.23 \ldots 0.70 \,\, {\rm fm}^2 \,\, ,
\eeq
where the bound is obtained by setting either $d_1=0$ or $d_2=0$ and
fixing the non--zero LEC from the proton value.
In what follows, we use $\langle r_A^2 \rangle_{p \to \Lambda \, K^+}
= 0.55\,$fm$^2$.
In the electroproduction case ($k^2 <0$), one is also sensitive to kaon
and hyperon electromagnetic form factors. For the $K^0 \, 
\Sigma^+$ final state,
the neutral kaon form factor to one loop is free of counter terms
and rather accurately given \cite{gl85}. The isovector Dirac radius
for the $\Sigma^+$ follows to be
\beq
\langle r^2_V \rangle_{\Sigma^+} =
 -\frac{(D+F)^2}{8\pi^2 F_K^2} \, \biggl( 5\ln\frac{M_K}{\lambda}
 - \frac{7}{2} \biggr)
 -\frac{ D^2+3F^2 }{12\pi^2 F_\pi^2} \, \biggl( 5 \ln\frac{M_\pi}{\lambda}
-\frac{7}{2} \biggr)    %%\nonumber \\
 +  12 \, d_{14}+4 \, d_{15}  \,\,\, .
\eeq
Remarkably, the same  combination of the LECs $d_{14}+d_{15}$
appears in the isovector proton radius. Taking the value of the most recent
dispersion--theoretical calculation for $\langle r^2_V \rangle_p 
=0.59\,$fm$^2$ \cite{MMD}, we predict
\beq
\langle r^2_V \rangle_{\Sigma^+} = 0.27 \ldots 0.33  
 \, \,{\rm fm}^2 \,\, ,
\eeq
where the smaller (larger) value refers  $F = 0.85 \, (0.75)$
and $D = 0.52 \, (0.50)$, respectively.
We now turn to the loop contributions. In charged kaon production, 
these lead to complex amplitudes due to the pion rescattering diagram
whose threshold is way below the physical production
threshold. Consequently, the imaginary parts calculated are
significantly too large. This can be understood in terms of the
Fermi--Watson theorem. Consider e.g. $\gamma \, p \to K^+ \, \Lambda$,
\beq
{\rm Im} \, E_{0+}^{{K^+}\Lambda} =  {\rm Re} \, E_{0+}^{\pi^+ n} \cdot
\biggl( -\frac{ \sqrt{3} \, M_K}{ \sqrt{8} \, F_\phi^2} \biggr)
 \cdot \frac{m_\Lambda}{4 \pi \sqrt{s}} \cdot {\rm PS} \,\,\,
\eeq
making use of the SU(3) generalization of the current algebra
prediction for the S--wave $\pi
N$ scattering lengths and PS denotes the pertinent phase space. 
If one now evaluates the electric dipole amplitude
at the $\pi^+ n$ threshold, $E_\gamma = 0.15\,$GeV, one exactly
reproduces the too large result of the one loop calculation to order $q^3$.
To get a more realistic value for these
imaginary parts, we take the amplitude $E_{0+}^{\pi^+ n}$ at the
energy corresponding to the $K \Lambda$ threshold, $E_\gamma =
0.91\,$GeV, from the recent compilation of the VPI group 
\cite{VPI}.\footnote{We neglect the momentum dependence of the P--waves
and the corrections to the strong scattering vertex. This needs to be 
improved.} This is similar to what was done in the $q^4$ calculation of
the process $\gamma p \to \pi^0 \pi^0 p$ \cite{bkm2pi0}. We remark
that these $q^3$ loop contributions are no longer finite as it is the
case in single pion production to this order (calculated in SU(2)). 
The renormalization
procedure to render these divergences finite is standard and discussed in
\cite{sven}.\footnote{Notice that the Feynman graph calculation can be
checked against the complete list of divergent operators in SU(3) and 
their $\beta$--functions given in ref.\cite{MM}.}
 In what follows, we use $\lambda =1\,$GeV, with $\lambda$
the scale of dimensional regularization.  Finally, to
estimate the LECs $d_3 , \ldots , d_{13}$, we use resonance saturation
including the baryon decuplet and the vector meson nonet. A detailed
account of this procedure can be found in ref.\cite{sven}

\bigskip

\noindent {\bf 4.} We now present the results for the various
final states (photoproduction case). We use the parameters listed
above. A more detailed account of the dependence on these parameters
is given in \cite{sven}.

\noindent ${\underline{K^0 \Sigma^+}:}$ All LECs are determined by
resonance exchange. In Fig.1a, we show the total cross section for the
first 100 MeV above threshold. No  data point exists in this range so
far, but soon the new ELSA data should be available \cite{schwille}. 
The electric dipole amplitude is real at threshold, we
have $E_{0+}^{\rm thr} (K^0 \Sigma^+) = 1.07 \times 10^{-3}/M_{\pi}$.
The coupled channel model based on a tree level chiral Lagrangian
at order $q^2$ gives a larger value
$E_{0+}^{\rm thr} (K^0 \Sigma^+) = (1.34 + 3.38i) \times
10^{-3}/M_{\pi}$ \cite{norb}. We remark that in that approach only
S--waves are considered and thus the electric dipole amplitude
effectively subsumes some of the P--wave contributions treated here.
In Fig.1b we show a prediction for the recoil polarization at
$E_\gamma = 1.26\,$GeV (which is the central energy of the lowest bin
of the not yet published ELSA data). 

\noindent ${\underline{K^+ \Lambda}:}$The total cross section from
threshold up to 100 MeV above is shown in Fig.2a. The lowest
$E_\gamma$ bin from ELSA \cite{bock} is $0.96 < E_\gamma < 1.01\,$GeV and has
$\sigma_{\rm tot} = (1.43 \pm 0.14)\, \mu$b, i.e. we slightly
underestimate the total cross section. In Fig.2b, we show the predicted
recoil polarization $P$ at $E_\gamma = 1.21\,$ GeV (which is higher in
energy than our approach is suited for). Amazingly, the shape and
magnitude of the data \cite{bock} is well described for forward angles,
but comes out on the small side for backward angles.  
Isobar models \cite{as}\cite{wjc}\cite{mbhw},
that give a descent description of the total and differential 
cross sections also at higher energies, fail to explain this angular
dependence of the recoil polarization (an exception is the recent work
of ref.\cite{st}).

\noindent ${\underline{K^+ \Sigma^0}:}$The total cross section is
shown in Fig.3a. It agrees  with the two data points from 
ELSA \cite{bock}. The recoil polarization at $E_\gamma =
1.26\,$GeV is shown in Fig.3b. It has the right shape but comes out
too small in magnitude. Nevertheless, we observe the important sign 
difference to the $K^+ \Lambda$ case, which is commonly attributed to 
the different quark spin structure of the $\Lambda$ and the 
$\Sigma^0$.\footnote{Notice that this argument is strictly correct
for massless quarks only.}
Here, it stems from an intricate interference of the complex S-- and
P--wave multipoles. In particular, one needs both s-- and t--channel
resonance excitations to get the shape of $P$.
In any case, one would like to have data closer to
threshold and with finer energy binning to really test the CHPT scheme.

\bigskip

\noindent {\bf 5.} In summary, we have used three--flavor chiral
perturbation theory to investigate threshold kaon production off
protons by real (and virtual) photons. Fixing all unknown parameters
from single baryon properties and making use of the resonance
saturation hypothesis, we achieve a satisfactory description of the
few threshold data. In particular, the shape of the angular dependences
of the $\Lambda$ and $\Sigma^0$ polarization at $E_\gamma \simeq
1.2\,$GeV is similar to the empirical ones. Clearly, these results 
should only be considered indicative since we have to include (a) higher 
order effects (for both the S-- and P--waves), (b) higher partial waves
and (c) have to  get a better handle
on the ranges of the various coupling constants. In addition, one would
also need more data closer to threshold, i.e. in a region where the
method is applicable. However, the results presented are encouraging 
enough  to pursue a more detailed study of these reactions (for real 
and virtual photons) in the framework of chiral perturbation theory.

\bigskip \bigskip
%%%%%%%%%%%%%%%%%% REFERENCES %%%%%%%%%%%%%%%%%%%%%%%%%%%%

%%%%%%%%%%%%%%%%%% Figures  %%%%%%%%%%%%%%%%%%%%%%%%%%%%
\bigskip \bigskip \bigskip

\section*{Figure Captions}

\begin{enumerate}

\item[Fig.~1] (a) Total cross section for $\gamma p \to K^0 \Sigma^+$
(solid line). The S--wave contribution is given by the dotted line.
(b) Recoil polarization at $E_\gamma = 1.26 \,$GeV. 
 
\item[Fig.~2] (a) Total cross section for $\gamma p \to K^+ \Lambda$
(solid line). The S--wave contribution is given by the dotted line.
(b) Recoil polarization at $E_\gamma = 1.21 \,$GeV. 
The data are from ref.\cite{bock}. The two data points from 
ref.\cite{erbe} in this interval are not shown.

\item[Fig.~2] (a) Total cross section for $\gamma p \to K^+ \Lambda$
(solid line). The S--wave contribution is given by the dotted line.
(b) Recoil polarization at $E_\gamma = 1.26 \,$GeV. 
The data are from ref.\cite{bock}. 

\end{enumerate}

\newpage

\vskip -3cm

\epsfysize=8.6in
\epsffile{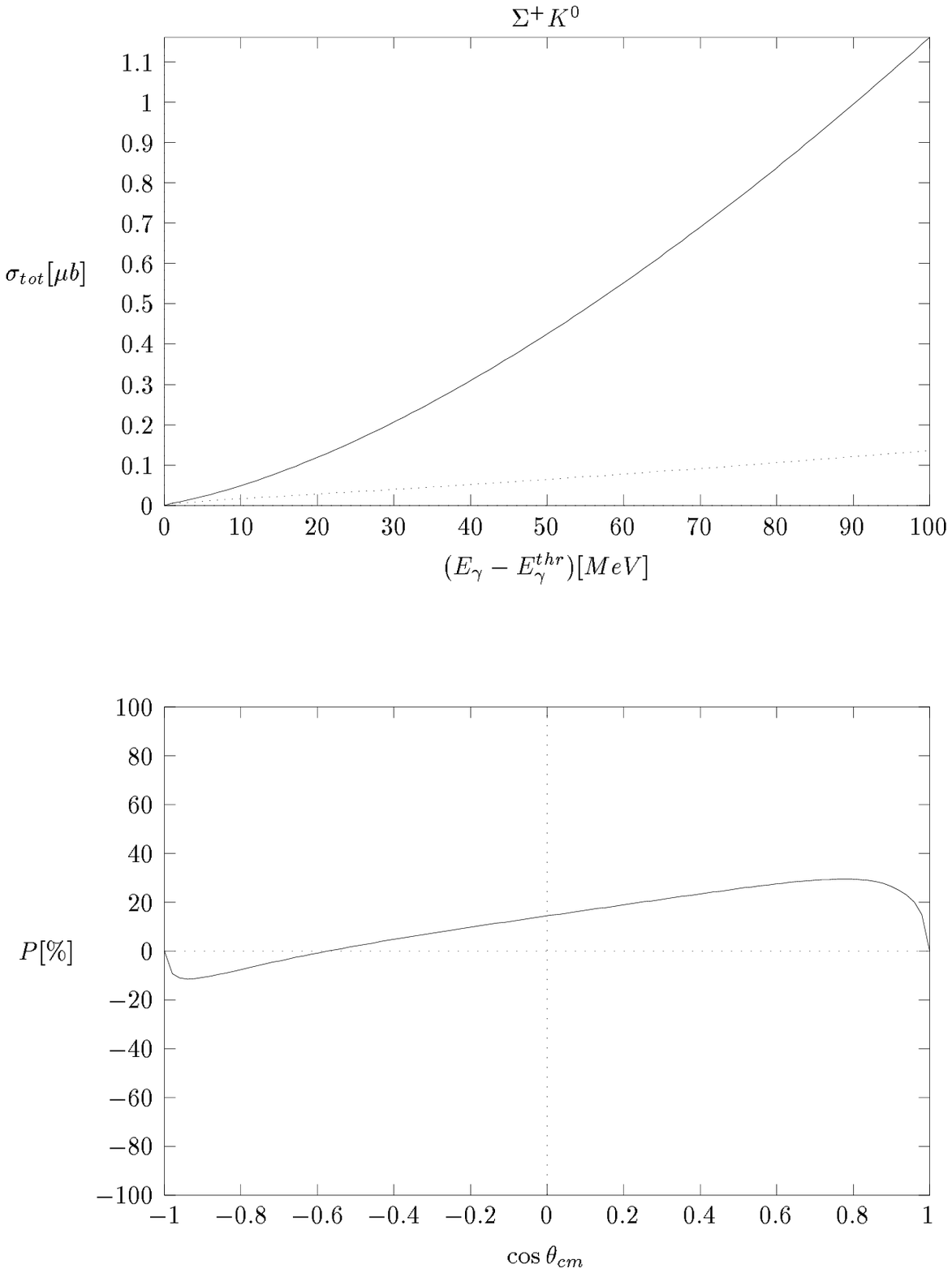}

\vskip -2cm

\centerline{\Large Figure 1}

\newpage

\vskip -3cm

\epsfysize=8.6in
\epsffile{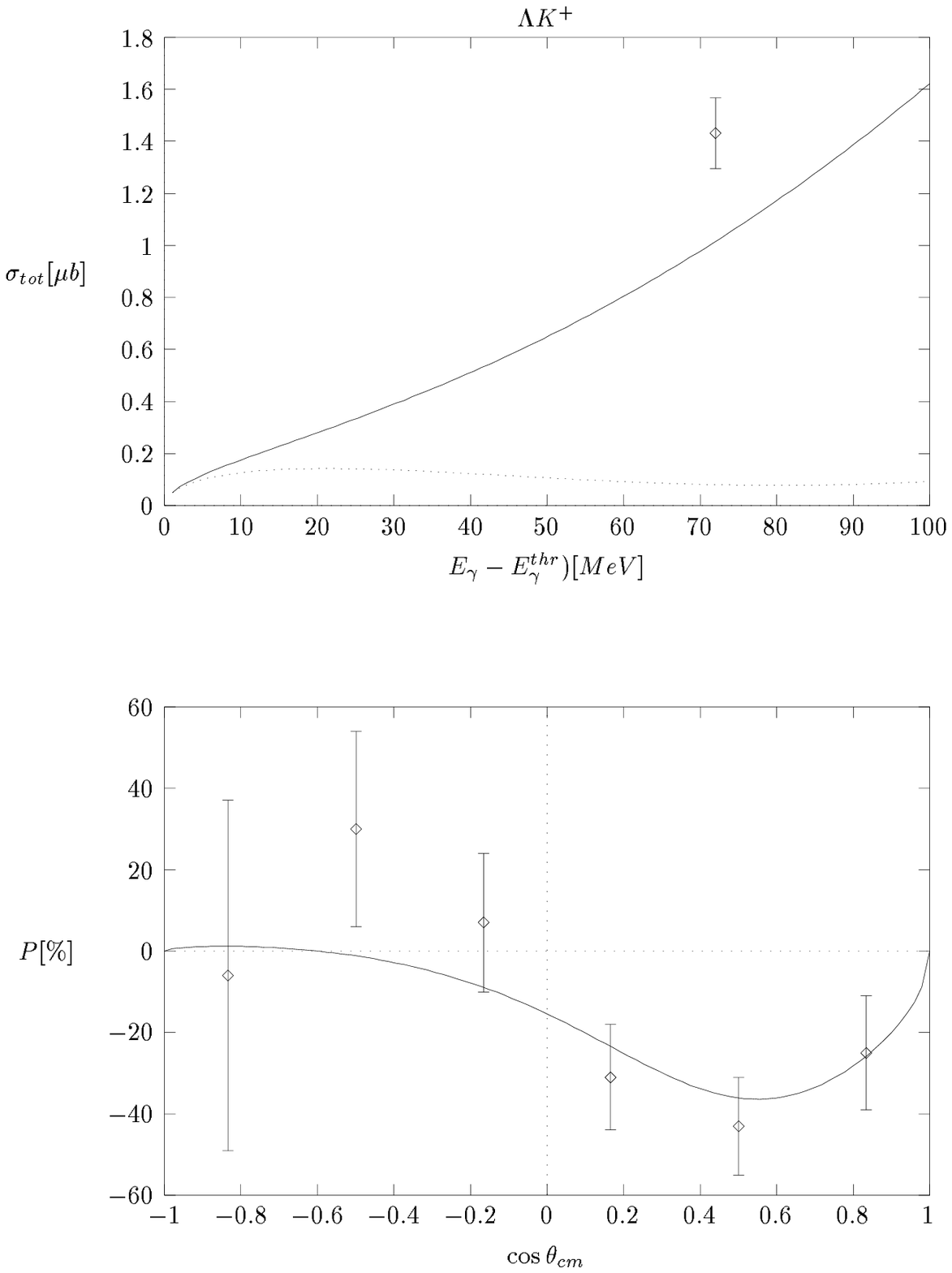}

\vskip -2cm

\centerline{\Large Figure 2}

\newpage

\vskip -3cm

\epsfysize=8.6in
\epsffile{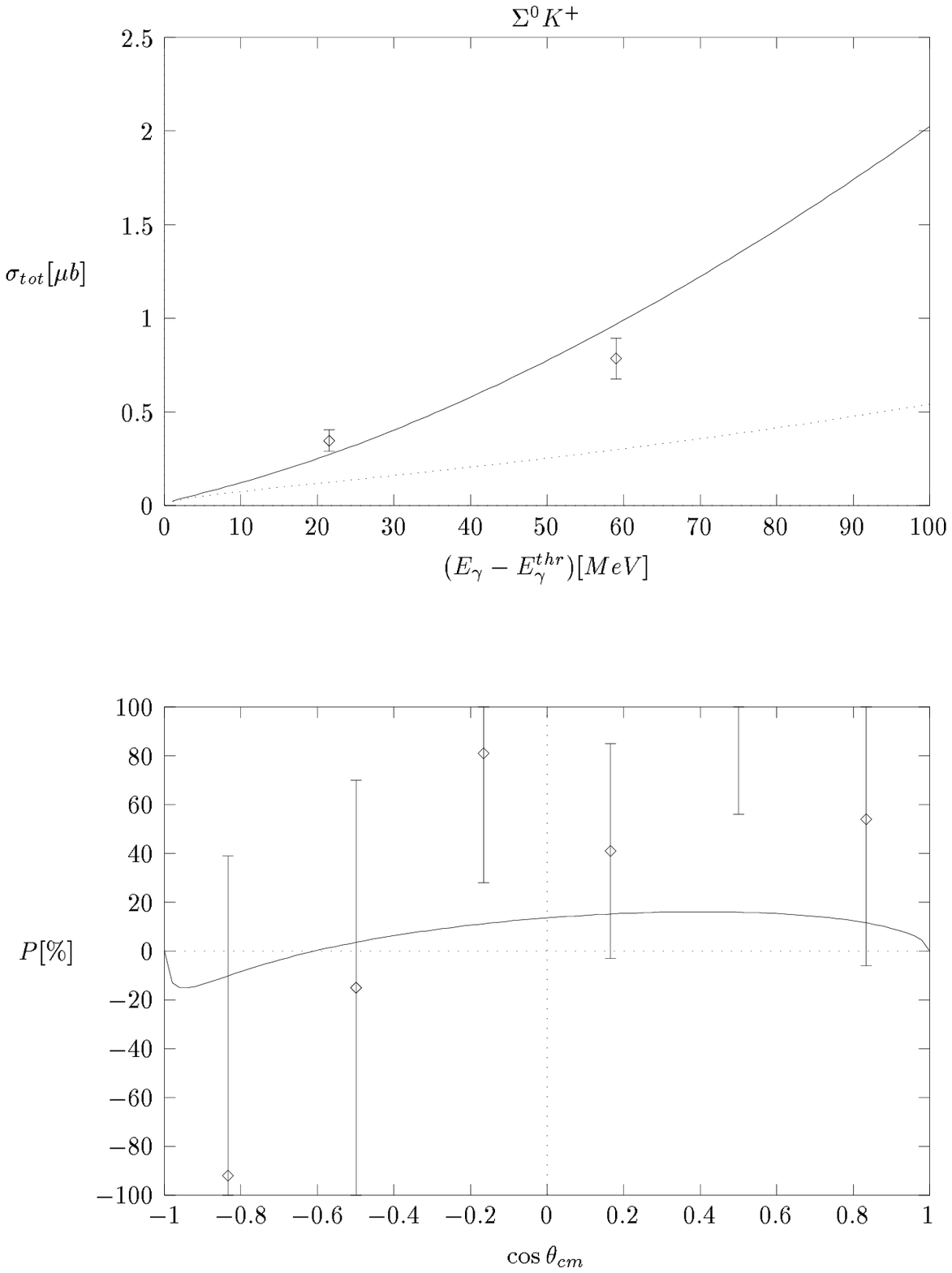}

\vskip -2cm

\centerline{\Large Figure 3}

\end{document}